# Amplification of light pulses with orbital angular momentum (OAM) in nitrogen ions lasing

Haicheng Mei[1,†], Jingsong Gao[2,†], Kailu Wang[1], Jiahao Dong[1], Qihuang Gong[2], Chengyin Wu[2], Yunquan Liu[2], Hongbing Jiang[2,#], and Yi Liu[1,3,*]

[1] *Shanghai Key Lab of Modern Optical System, University of Shanghai for Science and Technology, 516, Jungong Road, 200093 Shanghai, China*
[2] *State Key Laboratory for Mesoscopic Physics, School of Physics, Peking University, Beijing 100871, China*
[3] *CAS Center for Excellence in Ultra-intense Laser Science, Shanghai, 201800, China*
[†]*These authors contributed equally to this work.*

*\*yi.liu@usst.edu.cn, #hbjiang@pku.edu.cn*

**Abstract:** Nitrogen ions pumped by intense femtosecond laser pulses give rise to optical amplification in the ultraviolet range. Here, we demonstrated that a seed light pulse carrying orbital angular momentum (OAM) can be significantly amplified in nitrogen plasma excited by a Gaussian femtosecond laser pulse. With the topological charge of $\ell = \pm 1$, we observed an energy amplification of the seed light pulse by two orders of magnitude, while the amplified pulse carries the same OAM as the incident seed pulse. Moreover, we show that a spatial misalignment of the plasma amplifier with the OAM seed beam leads to an amplified emission of Gaussian mode without OAM, due to the special spatial profile of the OAM seed pulse that presents a donut-shaped intensity distribution. Utilizing this misalignment, we can implement an optical switch that toggles the output signal between Gaussian mode and OAM mode. This work not only certifies the phase transfer from the seed light to the amplified signal, but also highlights the important role of spatial overlap of the donut-shaped seed beam with the gain region of the nitrogen plasma for the achievement of OAM beam amplification.

## 1. Introduction

Cavity-free lasing of air molecules or their derivative under excitation of intense ultrafast laser pulses are fascinating phenomena and has attracted many attentions in recent 10 years [1–11]. With properly chosen wavelengths of the pump laser, it has been demonstrated that many species of the air plasma, including $N_2^+$, $N_2$, O atom, $CO_2^+$, even water vapor, can serve as gain media and produce coherent optical emission in the forward and/or the backward directions [6–9,12–15]. Air lasing is interesting from the perspective of fundamental research due to the rich high-field effects involved. In the mean time, it holds unique potential for optical remote sensing since it can provide a coherent optical beam from the sky to the ground observer. Among the different kinds of air lasing effects, the lasing of $N_2^+$ has obtained most of the attentions [16–19]. This is not only due to the abundance of $N_2$ in air, but also to the fact that these lasing effects can be activated by the intense 800 nm femtosecond pulses delivered by Ti: Sapphire laser system [3,20–23], which is the most widely available laser in the ultrafast optics labs. Under the excitation of 800 nm femtosecond laser pulses, nitrogen ions emit coherent 391.4 and 427.8 nm emission in the forward direction. With the injection of external seed pulse around these two wavelengths, it has been widely observed that significant optical amplification by 2-3 orders of magnitude can be obtained [24]. These two wavelengths correspond to the optical transitions of the second excited states of nitrogen ions $B^2\Sigma_u^+(v'=0)$ to the ground state $X^2\Sigma_g^+(v=0,1)$, where $v'$ and $v$ denote the vibrational quantum number of the corresponding states. Although the mechanism for the optical amplification for $N_2^+$ lasing is still under debated, its nature has been accepted as superfluorescence [25–27]. While in most of the previous studies linearly polarized 800 nm femtosecond pulses were employed, several other forms of pump laser fields, including two color counter-rotating fields [28], polarization screwed optical pulse [29], have been used to clarify the physical mechanism or to improve the conversion efficiency. Recently, structured light has attracted great interest [30–32]. In 2023, the concept of orbital angular momentum (OAM) was introduced to the study of $N_2^+$ lasing [33,34].

Some of the current authors have demonstrated, for the first time, that a vector or vortex pump pulse can generate an amplified 391.4 nm emission with vector or vortex properties [33]. In particular, it was shown that a femtosecond vortex pump pulse with a topological charge of ℓ = 1 can directly generate a vortex beam at 391.4 nm carrying OAM with ℓ = 1, and can also amplify an external Gaussian seed beam into an vortex beam with ℓ = 1. This opens a new degree of freedom for the study of $N_2^+$ air lasing and holds unique potential for the remote generation and amplification of OAM beams in ambient air. Soon later, Y. Hu *et al.* also employed a vortex pump or seed pulse to produce $N_2^+$ lasing effect [34]. They claimed that an OAM seed pulse with ℓ = 2 was amplified into a conventional Gaussian beam, and the OAM information of the seed pulse was wiped off in the amplification process. This is surprising since one would expect that the spatial phase information of the seeding pulse will be encoded in the amplification process and an amplified emission with OAM should be expected.

In this work, we employed a Gaussian femtosecond laser pulse to excite a nitrogen gas plasma and injected a seed pulse with OAM of ℓ = ± 1. It was observed that the seed pulse can be significantly amplified by about 1000 times, and the amplified emission inherits the topological charge of the incident seed pulse, in contrast with the recent report by Y. Hu *et al* where they claimed that the OAM seed pulse was amplified with the OAM property lost. We further show that the spatial alignment of the pump laser-induced plasma with the seed beam plays an important role, *i.e.*, a misalignment leads to degenaracy of the amplified OAM beam to a Gaussian beam. This work opens the avenue of OAM beam amplification in $N_2^+$ air lasing.

## 2. Experimental setup

The schematic of the experiment is depicted in Fig. 1. A commercial Ti: Sapphire femtosecond laser system (Legend-Elite-Duo HE+, Coherent Ltd) was employed in the experiments. The laser system produces 35 fs pulses at a central wavelength of 796 nm, at a repetition rate of 1 kHz. The output femtosecond laser pulses were split into two replica, one pulse serves as the pump pulse with a pulse energy of ~3 mJ, and the other pass through a BBO crystal to generate its second harmonic around 398 nm. A spiral phase plate (SPP) was installed on the seed beam path to obtain an OAM seed pulse with the topological charge of $\ell = \pm 1$. As shown in the inset of Fig. 1, the beam profiles of the pump and seed pulses are Gaussian and donut-shaped, respectively, which are recorded by a charge-coupled device (CCD). The Gaussian pump pulse and the OAM seed pulse are collinearly combined by a dichroic mirror (DM) with high reflectivity around 400 nm and high transmission around 800 nm. They are then focused by a lens (L1) with a focal length of 30 cm into the gas chamber filled with 20 mbar nitrogen gas. After the chamber, the lasing emission from the nitrogen gas plasma is collimated by a lens (L2), spectrally filtered by a shortpass filter at 650 nm and two narrow bandpass filters at 390 ± 5 nm. The emission signal is recorded by a fiber spectroscopy or a CCD.

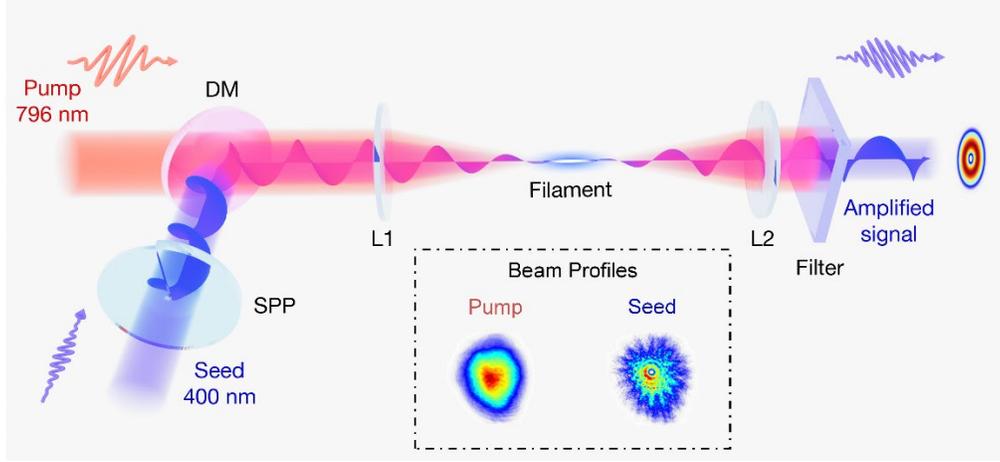

Fig. 1. Schematic diagram of the experimental setup. The seed pulse around 400 nm with the topological charge of $\ell = 1$ was generated by a spiral phase plate (SPP). Inset: The spatial profile of the pump and seed beams.

## 3. Results and discussion

First, the vortex seed beam with the topological charge of $\ell = 1$ is employed. The spectra of the emissions are shown in Fig. 2. The signal at 391 nm is associated with the transition of $B^2\Sigma_u^+(v' = 0) \rightarrow X^2\Sigma_g^+(v = 0)$. When nitrogen gas is pumped by only the pump pulse, there is a very weak self-seeded signal [4]. With the injection of the OAM seed pulse at an optimal delay, the emission energy is significantly enhanced by about 2-3 orders of magnitude. The intensity of the amplified signal is around 900 times as that of the seed beam. The corresponding spatial beam profiles of the seed and the amplified signal are shown in Fig. 3(a) and (c) . The beam profile of the seed pulse with an intensity singularity is the typical shape generated by the SPP, and we can notice the boundaries between the neighboring areas because the thickness of the SPP is not continuously varied. One notes that the amplified signal is donut-shaped and also have an intensity singularity, indicating that it is an OAM beam. The emission intensity changes with the time delay between the pump and seed pulses, keeping the donut-shape unaltered. To measure the topological charge of the amplified pulses, we focus it by a cylindrical lens. The image at the focus is shown in Fig. 3(d). It is evident that the topological charge of the amplified signal is 1, identical to that of the seed beam whose focal image is shown in Fig. 3(b). Then we set the topological charge of the seed to be -1 by turning the back surfaces of the SPP to the front, presented in Fig. 3(f). The corresponding spatial beam profiles of the seed and the amplified signal are shown in Fig. 3(e) and (g). Their cylindrically focused images show that the topological charge of the amplified signal becomes $\ell$ = -1, as shown in Fig. 3(h).

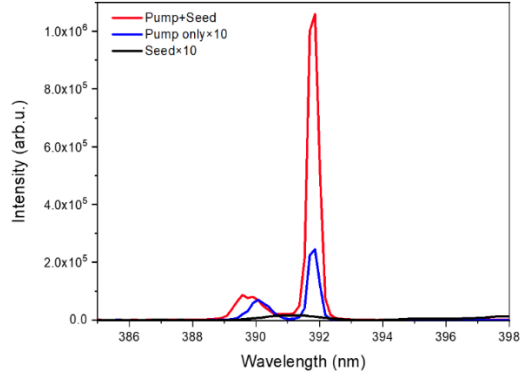

Fig. 2. The spectra of the amplified OAM pulse (red line), the seed pulse (black line), and the self-seeded 391.4 nm emission produced by only the pump pulse (blue line).

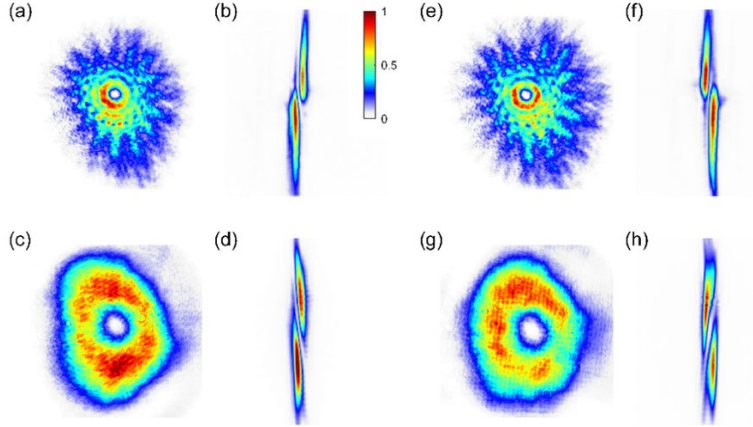

Fig. 3 Beam profile and corresponding cylindrically focused images of the seed and the amplified signal. (a)-(b), seed beam with $\ell$ = 1 and the image of its focus. (c)-(d), the corresponding amplified signal and the image of the focus. In (e)-(f) and (g)-(h), the topological charge is $\ell$ = - 1.

The above experimental results suggest that the spatial helical phase structure of the seed beam is recorded in the amplified emission signal. This observation is in accordance with theoretical expectations, whether in traditional stimulated emission or a superfluorescence process [35–37]. That is, the spatial phase of the amplified signal originates from the seed beam in a point-to-point manner in space, which means that the helical phase relationship between different spatial points in the seed beam is maintained in the amplified signal. To be specific, with the injection of the vortex seed pulse, the slowly varying part of the induced dipole moment of a nitrogen ion is given by [35,36]

$$p(x,y,z) = -\frac{i}{\hbar\gamma}d_{BX}^2 E(x,y,z), \quad (1)$$

where $\hbar$ is the reduced Planck constant, $\gamma$ is the decay rate from the B to the X state, $d_{BX}$ is the transition dipole moment between the B and X states of $N_2^+$, $E(x,y,z)$ is the electric field of the seed pulse in the gain medium. Because the pump and seed light fields are both x-polarized, all equations are expressed in the scalar form. We assume the injected seed light field is $E(x, y, z = 0) = E_0 e^{i\ell\phi}$, where $e^{i\ell\phi}$ is Hilbert factor possessing the spiral helical phase structure. The gain medium is a large number of ionized nitrogen molecules. The slowly varying amplitude of the macroscopic polarization of the gain medium is expressed as

$$P(x, y, z) = Np = -\frac{i}{\hbar\gamma} N(x, y) d_{BX}^2 E(x, y, z), \quad (2)$$

where $N(x, y)$ represents the average inversion number of nitrogen ions per unit volume, which is determined by the cross section of the pump beam. If the pump beam profile is donut-shaped, $N(x, y)$ will be donut-shaped too. To make the physical relationship clear, here we only consider the propagation term and neglect the diffraction effect. Thus, the wave equation (under the condition of a slowly varying approximation) is expressed as

$$2ik \frac{\partial E(x, y, z)}{\partial z} = -\mu\omega^2 P(x, y, z), \quad (3)$$

where $k$ is the wavenumber, $\mu$ is the permeability, $\omega$ is the angular frequency of the electric field. The Eq. (3) has an analytic solution, namely,

$$E(x, y, z) = E(x, y, z = 0)e^{Az} = E_0 e^{Az} e^{i\ell\phi}, \quad (4)$$

where $A$ represents the amplification coefficient and is expressed as $N(x, y)\mu\omega^2 d_{BX}^2 / 2k\hbar\gamma$. As the Eq. (1-4) present, the spatial helical phase term, $e^{i\ell\phi}$, naturally transfers from the seed beam $E(x, y, z = 0)$ to the dipole moment $p(x, y, z)$. The macroscopic polarization $P(x, y, z)$ takes this spatial phase information and finally it is transferred to the output signal $E(x, y, z = L)$, in the amplification process.

In contrast, surprisingly, Y. Hu *et al.* claimed that a vortex seed beam with $\ell$ = 2 was amplified into a Gaussian beam when pumped by an 800 nm Gaussian light [34], leading to the conclusion that the spatial helical phase structure of the vortex seed beam was wiped off in the amplification process in $N_2^+$ lasing. Our experimental observations are in sharp contrast with the results by Y. Hu *et al*. We found that the spatial alignment of the pump and the seed pulses is crucial for the spatial mode of the output signal. The Gaussian pump pulse produces a thin plasma channel in nitrogen gas that serves as an optical amplifier with maximum gain at the transverse center. In contrast, the seed beam exhibits a donut-shaped profile, where the intensity is null at the center. Therefore, to achieve a uniform optical amplification of the vortex seed beam, the spatial overlap of the donut-shaped seed pulse with the thin plasma gain medium turns out to be important.

In the experiments, we intentionally tilted the DM in the left, right, up and down directions to introduce misalignment of the pump and seed pulses. As shown in Fig. 4, the output modes become to be Gaussian mode with topological charge of $\ell$ = 0, instead of the OAM mode shown in Fig. 3. The schematic diagrams of the spatial overlap between the seed and pump beams is showed at the upper right corner of each panel, in which the red solid circles and the blue annulus represent the pump and the seed beam, respectively. We also noticed that the amplification ratio here is larger than that in Fig. 3. This is because the portion with higher intensity of the donut-shaped seed beam overlaps with the center of gain region.

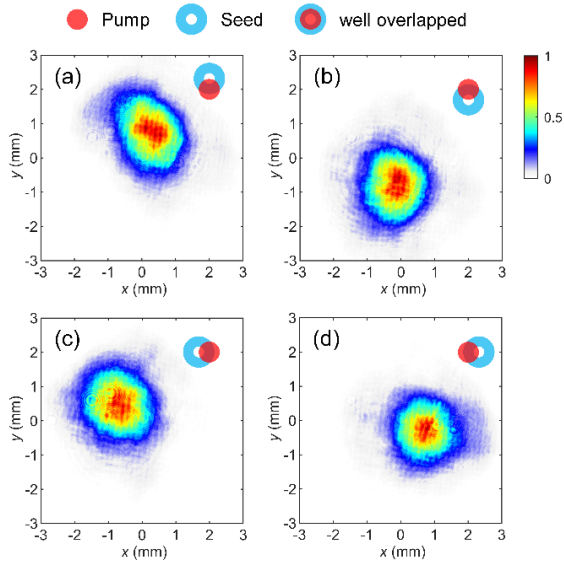

Fig. 4. Effect of misalignment of the focus of the OAM seed pulse with respect to the thin plasma amplifier created by the pump laser. In (a)-(d), the seed pulse was intentionally misaligned in the up, dowm, left, right directions, as illustrated by the insets. Insets are the illustrations of spatial overlap where the red solid circle represents pump beam and the blue annulus represents the seed beam.

To intuitively demonstrate the impact of spatial overlap on the results, we have recorded a video that dynamically captures how the tilting of the DM changes the spatial mode of the amplified signal. As shown in Fig. 5, we picked two frames of the video before and after adjusting the *y*-direction screw of DM. We strongly recommend the readers to watch the full video in the Supplementary Information, as it visually demonstrates this process.

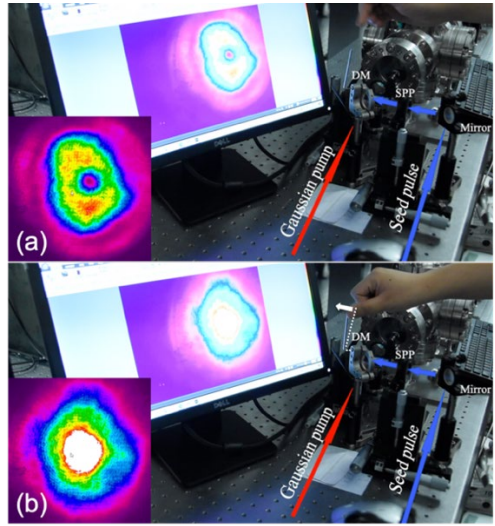

Fig. 5. Frames of the supplementary video correspond to (a) before and (b) after adjusting the *y*-direction screw of DM. See the full video in Supplementary Information. The bottom left illustration shows the images captured by the CCD and synchronized with the computer screen.

Instead of introducing misalignment of the pump and seed beam, we have also displaced the SPP, to spatially shift the position of the singular point of the seed beam while the alignment of the pump and seed beams remains untouched. The corresponding results are presented in Fig. 6. The displacement of the optical singularity on the seed beam is evident in Fig. 6(a-d), in contrast with those in Fig. 3(a) where the SPP is well aligned. We found that in these cases there are always positions to obtain amplified emission signal with good Gaussian modes, as shown in Fig. 6(e-h). Therefore, it becomes clear that the spatial alignment of the pump and OAM seed pulses (Fig. 4) and a spatially well positioned phase singularity of the seed beam (Fig. 6) are both crucial for the achievement of an amplified vortex emission. The effect of shifting the SPP on the output mode was also recorded in the video (see Supplementary Information). From a technical perspective, we can easily switch back and forth between Gaussian mode and vortex mode in such a cavity-free amplifier, which can be employed as an optical switch of OAM.

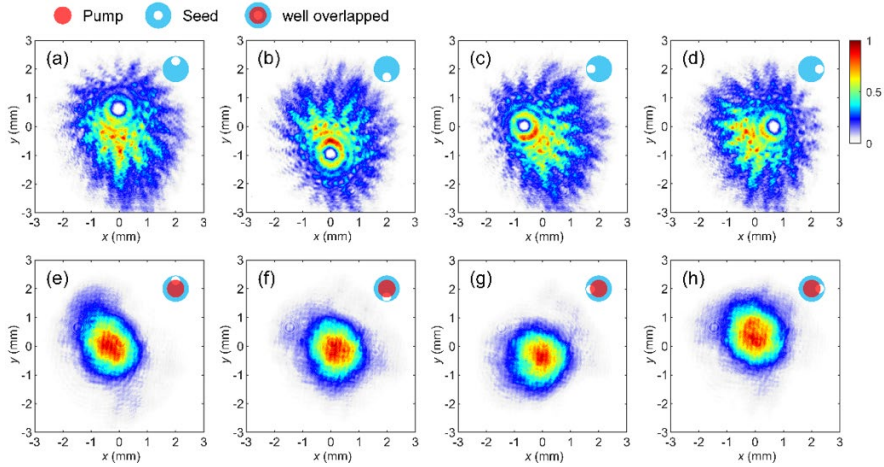

Fig. 6 Profiles of the seed pulse (a-d) and amplified emission (e-h) when the SPP is shifted to up, down, left and right sides. Insets are the illustrations of spatial overlap where the red solid circle represents pump beam and the blue annulus represents the seed beam.

To get insight into the effect of the misalignment of the Gaussian pump and vortex seed beams, we have employed numerical simulations to study the evolution of both the intensity and phase distribution of the OAM seed beam during amplification along propagation. Because the evolution of the spatial mode is concerned, the diffraction effect presented by the Laplacian term should now be considered in the wave equation. Thus, Eq. (3) becomes

$$\frac{\partial^2 E(x,y,z)}{\partial x^2} + \frac{\partial^2 E(x,y,z)}{\partial y^2} + 2ik\frac{\partial E(x,y,z)}{\partial z} = -\mu\omega^2 P(x,y,z). \tag{5}$$

We set the diameter of the Gaussian gain region, $N(x,y)$, to be 30 mm. The maximum gain is at the origin (0, 0). The light field of the injected seed pulse is the Laguerre-Gaussian mode ($\ell = 1$, p = 0) with a beam waist of 15 mm [38]. The corresponding numerical results are presented in Fig. 7. In Fig. 7(a), it is seen that a well-aligned vortex seed pulse is amplified while its donut-shaped intensity distribution and spatial spiral phase structure are kept. Under our parameter settings, the amplification ratio at z = 0, 2.5, 5, 7.5 and 10 mm is about 1, 6, 34, 170 and 689, respectively. The twisted spatial helical phase structure at z = 10 mm is caused by the propagation and diffraction effect. For the case of misalignment as shown in Fig. 7 (b), partial overlap of the seed beam with the gain media leads to a Gaussian intensity distribution after amplification and propagation. This is because only the portion of the donut-shaped seed beam covered by the gain region is selected and amplified. The amplification ratio at z = 0, 2.5, 5, 7.5 and 10 mm is about 1, 6, 48, 414 and 3812, respectively. The amplification ratio in case of misalignment is higher than that of well-aligned one, which is consistent with the experiment results. Due to the propagation effect, the phase of the gain-covered portion gradually becomes identical.

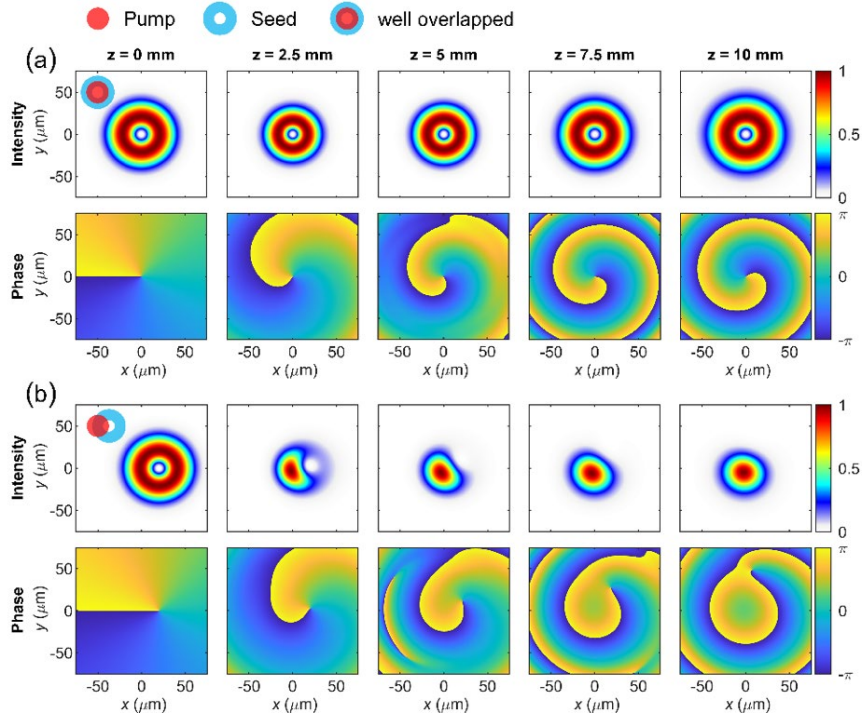

Fig. 7. Numerical simulation results of the OAM seed beam along propagation in plasma gain media of 10 mm. In (a) the seed beam is well aligned with the pump, while in (b) the two beams are misaligned in the horizontal direction. In both (a) and (b), the upper row show the intensity distribution and the lower row corresponds to the spatial phase distribution. The amplification ratio at z = 0 mm, 2.5 mm, 5 mm, 7.5 mm and 10 mm is about (a) 1, 6, 34, 170 and 689, respectively; (b) 1, 6, 48, 414 and 3812, respectively. Insets are the illustrations of spatial overlap where the red solid circle represents pump beam and the blue annulus represents the seed beam.

As a final remark, we would like to point out that the intensity of the vortex seed pulse is not so crucial for the observation of an amplified OAM beam. In our experiments, we have reduced the intensity of the vortex seed light to the detection limit of the spectrometer while we increased the integration time of the spectrometer by 2000 times in comparison to that of the 391 nm lasing. We found the output signal is also an amplified vortex beam, whose intensity is ~12000 times as that of the attenuated seed light (see the video in Supplementary Information). This means that the helical phase structure of the seed beam is always maintained in the output signal, even though its input intensity is very weak. These observations support the statement in our recent work [33]: when the pump light is a vortex beam with $\ell = 1$, the fully extinguishing of the parasitic second harmonic of the vortex pump light generated by the waveplates is very necessary. The parasitic second harmonic inherently overlaps well with the pump light unless the incident angle of the pump light on the waveplates is large. Otherwise, the parasitic second harmonic at 400 nm carrying 2 units of OAM will play the role of an external seed pulse and then definitely contaminate the amplification process, even if it is very weak.

## 4. Conclusion

In conclusion, we have shown that a weak 400 nm seed pulse carrying OAM of $\ell = \pm 1$ can be substantially amplified in the nitrogen gas plasma excited by an intense 800 nm femtosecond Gaussian pulse, with the OAM property well maintained. This is different from a recent report that an OAM seed pulse was amplified into a conventional Gaussian beam in the amplification process. We found that the misalignment of the pump and seed pulse at the common focus can lead to the degeneracy of the amplified OAM beam ($\ell = \pm 1$) into a conventional Gaussian beam ($\ell$ = 0). This sensitive dependence of the OAM amplification should be due to the particular spatial intensity and phase distribution of the seed pulse at the focus, where it manifests as a donut-shaped profile with an inner singular spot. Consequently, a misalignment of the donut-shaped seed pulse with the pump-induced plasma can break the cylindrical symmetry and leads to amplification into a Gaussian beam. From a technical perspective, we can easily switch the output signal back and forth between Gaussian mode and vortex mode in such a lasing system by simply shifting the SPP. This study unambiguously shows that seed pulse with OAM can be amplified in the nitrogen gas plasma via a cavity-free process and opens the avenue of OAM beam amplification in remote atmosphere, which can be interesting for free-space optical communication, remote sensing, as well as quantum information relay.

**Funding.** The work is supported in part by the National Natural Science Foundation of China (Grant Nos. 12034013, 11904232), and the National Key R&D Program of China (Grant No. 2018YFB2200401).

**Disclosures.** The authors declare no conflicts of interest.

**Data availability.** Data underlying the results presented in this paper are not publicly available at this time but may be obtained from the authors upon reasonable request.

**Supplemental document.** See Supplement Information for supporting content.